\documentclass[conference]{IEEEtran}
\IEEEoverridecommandlockouts
\usepackage{cite}
\usepackage{amsmath,amssymb,amsfonts}
\usepackage{algorithmic}
\usepackage{graphicx}
\usepackage{tikz}
\usepackage{pgfplots}
\pgfplotsset{compat=1.17}
\usepgfplotslibrary{groupplots}
\usepgfplotslibrary{fillbetween}
\usepackage{textcomp}
\usepackage{xcolor}
\usepackage{hyperref}
\usepackage{caption}
\usepackage{subcaption}
\def\BibTeX{{\rm B\kern-.05em{\sc i\kern-.025em b}\kern-.08em
    T\kern-.1667em\lower.7ex\hbox{E}\kern-.125emX}}

\newcommand{\mat}[1]{\mathbf{\MakeUppercase{#1}}}
\newcommand{\vect}[1]{\mathbf{\MakeLowercase{#1}}}

\newcommand{\R}{\mathbb{R}}

\usepackage{xspace}
\usepackage[toc, acronym, nogroupskip]{glossaries}
\usepackage{glossary-superragged}
\newacronym{DNN}{DNN}{Deep Neural Network}
\newacronym{LDA}{LDA}{Latent Dirichlet Allocation}
\newacronym{TCGA}{TCGA}{The Cancer Genome Atlas}
\newacronym{IG}{IG}{Integrated Gradient}
\newacronym{PGI}{PGI}{Prediction Gap on Important features}
\newacronym{PGU}{PGU}{Prediction Gap on Unimportant features}
\newacronym{PCR}{PCR}{Prediction Curve on Random features}
\newacronym{LR}{LR}{Logistic Regression}
\newacronym{MLP}{MLP}{MultiLayer Perceptron}
\newacronym{FA}{FA}{Feature Agreement}

\newcommand{\LDA}{\gls{LDA}\xspace}
\newcommand{\TCGA}{\gls{TCGA}\xspace}

\newcommand{\PGI}{\gls{PGI}\xspace}
\newcommand{\PGU}{\gls{PGU}\xspace}

\newcommand{\MLP}{\gls{MLP}\xspace}
\newcommand{\LR}{\gls{LR}\xspace}
\newcommand{\FA}{\gls{FA}\xspace}

\begin{document}

\title{
Studying Limits of Explainability by Integrated Gradients for Gene Expression Models
\thanks{This work was supported by the CHIST-ERA grant CHIST-ERA-19-XAI-006, for the GRAPHNEX ANR-21-CHR4-0009 project.
The results are partially based upon data from the TCGA Research Network: \url{https://www.cancer.gov/tcga}. We thank Maroun Bou Sleiman for his expertise on genomic data and all members of the GraphNEx team for the fruitful discussions.}
}

\author{\IEEEauthorblockN{Myriam Bontonou}
\IEEEauthorblockA{
\textit{Univ Lyon, ENSL, CNRS, LBMC}\\
Lyon, France \\
myriam.bontonou@ens-lyon.fr}
\and
\IEEEauthorblockN{Anaïs Haget}
\IEEEauthorblockA{
\textit{LTS2 laboratory, EPFL}\\
Lausanne, Switzerland \\
anais.haget@epfl.ch}
\and
\IEEEauthorblockN{Maria Boulougouri}
\IEEEauthorblockA{
\textit{LTS2 laboratory, EPFL}\\
Lausanne, Switzerland \\
maria.boulougouri@epfl.ch}
\and
\IEEEauthorblockN{Jean-Michel Arbona}
\IEEEauthorblockA{
\textit{Univ Lyon, ENSL, CNRS, LBMC}\\
Lyon, France \\
jeanmichel.arbona@ens-lyon.fr}
\and
\IEEEauthorblockN{Benjamin Audit}
\IEEEauthorblockA{
\textit{Univ Lyon, ENS de Lyon, CNRS, Laboratoire de physique}\\
Lyon, France \\
benjamin.audit@ens-lyon.fr}
\and
\IEEEauthorblockN{Pierre Borgnat}
\IEEEauthorblockA{
\textit{Univ Lyon, ENS de Lyon, CNRS, Laboratoire de physique}\\
Lyon, France \\
pierre.borgnat@ens-lyon.fr}
}

\maketitle

\begin{abstract}
Understanding the molecular processes that drive cellular life is a fundamental question in biological research. Ambitious programs have gathered a number of molecular datasets on large populations. To decipher the complex cellular interactions, recent work has turned to supervised machine learning methods. The scientific questions are formulated as classical learning problems on tabular data or on graphs, e.g.\ phenotype prediction from gene expression data. In these works, the input features on which the individual predictions are predominantly based are often interpreted as indicative of the cause of the phenotype, such as cancer identification.
Here, we propose to explore the relevance of the biomarkers identified by Integrated Gradients, an explainability method for feature attribution in machine learning. Through a motivating example on The Cancer Genome Atlas, we show that ranking features by importance is not enough to robustly identify biomarkers. As it is difficult to evaluate whether biomarkers reflect relevant causes without known ground truth, we simulate gene expression data by proposing a hierarchical model based on Latent Dirichlet Allocation models. We also highlight good practices for evaluating explanations for genomics data and propose a direction to derive more insights from these explanations.
\end{abstract}


\section{Introduction}
Understanding the molecular mechanisms that drive cellular metabolism is vital to better diagnose, treat and prevent diseases such as cancers or dementia.
Schematically, DNA encodes genes (genome) which are transcribed into RNA (transcriptome) and then translated into proteins (proteome) that catalyse the complex molecular processes. The expression of genes is regulated in part by transcription factors and by epigenetic mechanisms (epigenome). It can also be modified by genetic mutations. A long standing objective is to seek relations between measurements of gene expression and phenotypes such as clinical features. 

Ambitious programs have gathered such molecular databases on disease patients and on the general population, e.g.\ \TCGA for cancer study (\url{https://www.cancer.gov/tcga}), ROSMAP~\cite{de2018multi} about dementia, and the UK Biobank with genetic information on $\sim 500000$  people (\url{https://www.ukbiobank.ac.uk}).
Data from these projects were analysed to seek relations between genetic variation, gene expression and clinical features. This has been done using statistical tests, or relying on clustering methods that group individual samples based on their molecular profiles; see the Review~\cite{tomczak2015review} about \TCGA.
To account for more complex biological relationships, recent works have turned to supervised machine learning methods to probe the same questions.

Hence, the question is formulated as
a classical learning problems on tabular data (possibly seen as on graphs): the prediction of interactions between genes ~\cite{du2019gene2vec} or between multi-omics modalities~\cite{jagtap2022branenet}, prognosis prediction from gene expression data~\cite{hao2018pasnet}, phenotype prediction from similar inputs~\cite{ramirez2020classification, bourgeais2021deep}, from single nucleotide change in DNA~\cite{demetci2021multi} or from multiple modalities~\cite{zitnik2019machine}.
Once trained, these models achieve reasonable performance, and the individual predictions can be explained by the input features on which they are predominantly based. These features are then often interpreted as biomarkers for the studied phenomenon, e.g. for cancer identification~\cite{bourgeais2021deep, ramirez2020classification}. To which extent are these biomarkers indicative of the molecular pathways responsible for the phenotypes?
Here we address this question in the context of explainability for data processing.

To achieve explainability, one should adapt the learning method to the data, to the task complexity and the expected explanations~\cite{rudin2022interpretable}.
In the present work, we will rely on additive feature attribution methods~\cite{lundberg2017unified}. 
Especially, the Integrated Gradients (IG) method~\cite{sundararajan2017axiomatic} is a widely used approach that satisfies the completeness axiom and is computationally efficient.
Explanation reliability can then be quantified by various metrics; e.g.\ faithfulness, stability and fairness have been developed for tabular data~\cite{agarwal2022openxai}.
Is it efficient to identify biomarkers in biological data using IG? 

To study this question and explore the limits of explainability in this context, we put forward three contributions:
\begin{itemize}
\item Using \TCGA gene expression dataset as a relevant example, we show that ranking features by importance is not enough to robustly identify biomarkers. Our point is that such features should be both sufficient and necessary for the predictions.
\item To this end, we propose to systematically 
evaluate two metrics on genomic data: the \PGU estimating the number of important features necessary for a single prediction, the \PGI estimating the number of important features sufficient for a single prediction.
\item Seeking confidence and control over the proposed explanations, we propose to adapt the \LDA model ~\cite{blei2003latent} to generate data that have similar properties to biological data (e.g., the hierarchical properties of expression pathways). We then evaluate the explanations obtained on this \LDA model thanks to the IG attribution method. 
\end{itemize}

The article is organised as follows: Section~\ref{sec:background} introduces the data, the learning methods, the attribution methods and the metrics used. Section~\ref{sec:example} details the study on pan-cancer classification based on transcriptomic data from \TCGA. It highlights the importance of systematically estimating several metrics, before reporting important features. Section~\ref{sec:model} proposes a \LDA based generative model, and we then evaluate the quality of the generated explanations. Concluding remarks are in Section~\ref{sec:ccl}.
The code and data are available publicly on {\small{\url{https://github.com/mbonto/XAI_for_genomics}}}.

\section{Background}
\label{sec:background}
The objective is to solve a classification task over $C$ classes for gene expression data. A data sample, $\vect{x}\in\R^F$, contains the expression of $F$ genes (features).
The dataset is a collection of cells of various classes (cancer types in \TCGA).
For the present work, we consider classical supervised methods, denoted by $f(\cdot)$: \LR or \MLP. Nonetheless, the methodology could be extended to any learning architecture, e.g.\ neural networks. 
A softmax function is applied before the output of the model which is $f(\vect{x})\in\R^C$. 
Working in a supervised context, the parameters are updated by gradient descent in order to minimise a loss function on a set of training examples. As the classes in \TCGA are unbalanced, learning is evaluated on a test set by the so-called \emph{balanced accuracy}, computed as the average of the recalls obtained on each class.

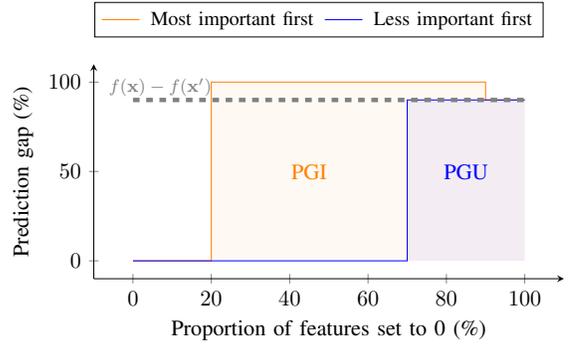
\begin{figure}[t]
\centering
\resizebox{0.85\linewidth}{!}{\begin{tikzpicture}
\begin{groupplot}[group style={group size=1 by 2, horizontal sep=1cm, vertical sep=0cm}]

    \nextgroupplot[
    hide axis,
    width=5cm,
    height=3cm,
    ymin=0,
    ymax=10,
    xmin=0,
    xmax=10,
    legend style={font=\small, at={(1.5,0.5)},anchor=east},
    legend columns=-1,
    column sep=0.1cm
    ]
    
    \addlegendimage{color=orange}
    \addlegendentry{Most important first}; 
    \addlegendimage{color=blue}
    \addlegendentry{Less important first};

    \nextgroupplot[
    width=\textwidth/2, 
    height=5cm,
    axis lines = left,
    enlargelimits = true,
    ylabel=Prediction gap (\%), 
    xlabel=Proportion of features set to 0 (\%), 
    ymin=0, 
    ymax=100,
    xmin=0,
    xmax=100,
    ylabel near ticks,]
    
    \addplot[name path=PGI, color=orange] table [x=Prop, y=PGI, col sep=comma] {Figures/curves.csv};
    \addplot[name path=PGU, color=blue] table [x=Prop, y=PGU, col sep=comma] {Figures/curves.csv};
    \addplot[color=gray, dashed, line width=2pt] table [x=Prop, y=Default, col sep=comma] {Figures/curves.csv};

    \path[name path=axis] (axis cs:0,0) -- (axis cs:100,0);
    \addplot [
        thick,
        color=orange,
        fill=orange, 
        fill opacity=0.05
    ]
    fill between[
        of=PGI and axis,
        soft clip={domain=0:100},
    ];

    \path[name path=axis] (axis cs:0,0) -- (axis cs:100,0);
    \addplot [
        thick,
        color=blue,
        fill=blue, 
        fill opacity=0.05
    ]
    fill between[
        of=PGU and axis,
        soft clip={domain=0:100},
    ];

    \node[blue] at (axis cs: 85, 50) {PGU};
    \node[orange] at (axis cs: 45, 50) {PGI};
    \node[gray] at (axis cs: 7, 97) {\footnotesize $f(\vect{x}) - f(\vect{x}^\prime)$};

\end{groupplot}
\end{tikzpicture}}
\caption{Scheme describing the Prediction Gaps on Important features (\PGI) and on Unimportant features (\PGU).}
\label{fig:pg}
\end{figure}

Explainability will be the ability of a method to propose biomarkers on a single prediction, i.e.\ which features are important for a prediction as estimated by a score $\phi_i$ computed for each feature $\vect{x}_i$. 
We choose the integrated gradients method (IG)~\cite{sundararajan2017axiomatic} rather than a perturbation based method such as SHAP~\cite{lundberg2017unified} because of its performance w.r.t. computation time.  To simplify the notations, $f(\vect{x})$ will denote the output associated to the true class $c$. Given a baseline $\vect{x}^\prime\in\R^F$, the IG score is:
\begin{equation}
    \phi_i(\vect{x}) = (\vect{x}_i - \vect{x}_i^\prime) \int_{\alpha=0}^1 \frac{\partial f(x)}{\partial \vect{x}_i} \bigg|_{x=\vect{x}^\prime + \alpha(\vect{x} - \vect{x}^\prime)}d\alpha\,.
\label{eq:IGdef}
\end{equation}
Integrated gradients satisfy the completeness property as the sum of the attributions is equal to the difference between the output of the network at the input and at the chosen baseline:
\begin{equation}
    \sum_{i=1}^F \phi_i(\vect{x}) = f(\vect{x}) - f(\vect{x}^\prime)\;.
\end{equation}
Note that a \LR is by itself interpretable as the weights' amplitudes reveal which features are important. The IG score of a feature used in a \LR simply reflects the product of the weight and the feature value. 

At a single prediction level, some metrics exist to evaluate the relevance of features~\cite{agarwal2022openxai}. \PGI and \PGU are two metrics measuring the faithfulness of an explanation. They are computed as the area under the curve of the prediction gap while a varying proportion of input features is set to 0 (the average value of features). Denoting $\vect{x}$ the original input and $\tilde{\vect{x}}$ the modified input (with some features set to 0), the prediction gap is: $\max (f(\vect{x}) - f(\tilde{\vect{x}}), 0)$. It increases as the features are removed. For \PGI (resp. \PGU), the features identified as the most (resp. less) important are removed first (Fig.~\ref{fig:pg}). By construction, the maximum area under the curves is 1. When important features are removed first, the model quickly makes wrong predictions, thus,  \PGI is expected close to 1. When less important features are removed first, the prediction stays stable until a large number of features is removed. When it occurs, the prediction of the model is close to the baseline prediction (zero input). As in practice, the observed area is restricted by $f(\vect{x}) - f(\vect{x}^\prime)$, the \PGU values can be adjusted for interpretation. Here, we divide them by $1 - f(\vect{x}^\prime)$. The prediction gaps are averaged over all test samples correctly classified. 

When the features that cause the class distinction are known, do they stand out among the most important features identified for the model as a whole? \FA is a metric that quantifies the number of a priori important features retrieved in the most important feature set identified. 
Denoting the set of features characteristic of a class as $\mathcal{F}$ and the set of features identified (here by IG) as the most important for the method as $\mathcal{M}$,
\begin{equation}
    \text{FA} = \frac{|\mathcal{F} \cap \mathcal{M}|}{|\mathcal{F}|}
\end{equation}
For instance, in our setting, the important features attributed to a class should be over-expressed genes resulting from over-expressed pathways characteristic of this class.

% \begin{figure}[b]
% \centerline{\includegraphics[width=0.8\linewidth]{Figures/class_imbalance.png}}
% \caption{Pan-Can \TCGA data: number of samples in each class}
% \label{fig:pan-can}
% \end{figure}

\section{Why ranking features by importance is not enough? A motivating example on Pan-Can TCGA}
\label{sec:example}

\subsection{Pan-Can TCGA dataset}

\TCGA, a cancer genomics program, generated genomic, epigenomic, transcriptomic, and proteomic datasets spanning 33 cancer types (publicly available at \url{https://portal.gdc.cancer.gov/}). In the present study, we consider a cancer type classification task on a gene expression dataset called \TCGA Pan-Can. It relies on 9680 samples classified in the 33 cancer types. Class size ranges from 36 to 1095 samples. For each sample, the expression of 16335 genes is measured.
The transcriptomic data have been pre-processed by the Pan-Cancer Atlas initiative to mitigate the bias induced by the diversity of the experimental settings~\cite{hoadley2018cell}.
The gene expression features are $\log_2(count+1)$ where $count$ is the upper-quartile normalized raw count, so as to reduce the impact of outliers.

\subsection{Learning method and experimental setting}
\label{subsec:exp_TCGA}
For classification learning, we consider a \LR and a \MLP. We also include a diffusion layer (D) beforehand; it diffuses the initial gene expression vector on the gene correlation graph. The relevance of this diffusion will be later motivated in Section~\ref{sec:model}. Before training, each feature (gene expression) is standardised. The attribution scores are computed by IG using a zero baseline in Eq.~(\ref{eq:IGdef}). The code uses Pytorch~\cite{paszke2019pytorch} and the Captum library~\cite{kokhlikyan2020captum}. More details are in Appendix~\ref{app:2}.

\subsection{Interpretation}
The results are presented in Table~\ref{tab:TCGA}. The balanced accuracy is slightly higher for the \MLP than for the \LR. The diffusion does not significantly degrade the performance. For the \LR, a PGU of 0.003 can be interpreted as the possibility to remove every gene except the 49 most important ones from the model without affecting its prediction. A PGI of 0.95 implies that the 817 most important genes should be removed from the model to hurt the prediction. 
Hence, the 49 genes are sufficient (PGU) but not necessary (PGI) to get a correct prediction. For the \MLP, even more genes could be removed without hurting the performance. Similar results are obtained for the diffused models.
This probably underlies the biological complexity.

\begin{figure}[t]
\centering
\begin{subfigure}[b]{\linewidth}
\caption{Pan-Can TCGA}
\centerline{\includegraphics[width=\linewidth]{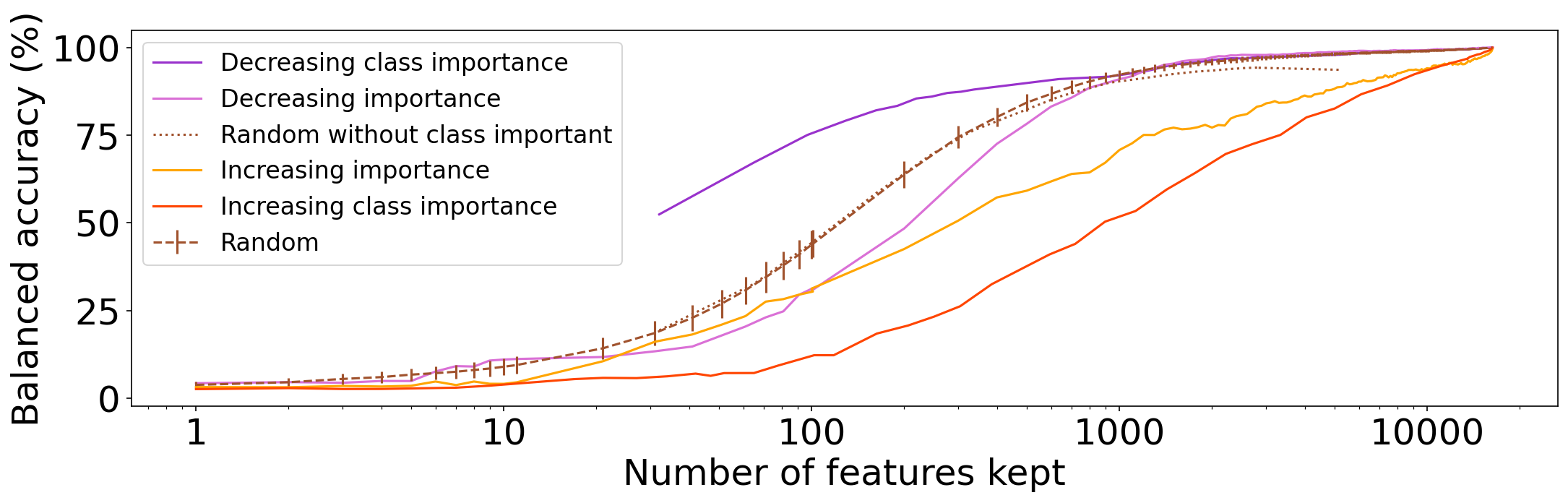}}
\label{fig:TCGA}
\label{fig:pancan_global}
\end{subfigure}

 \begin{subfigure}[b]{\linewidth}
 \caption{SIMU1}
\centerline{\includegraphics[width=\linewidth]{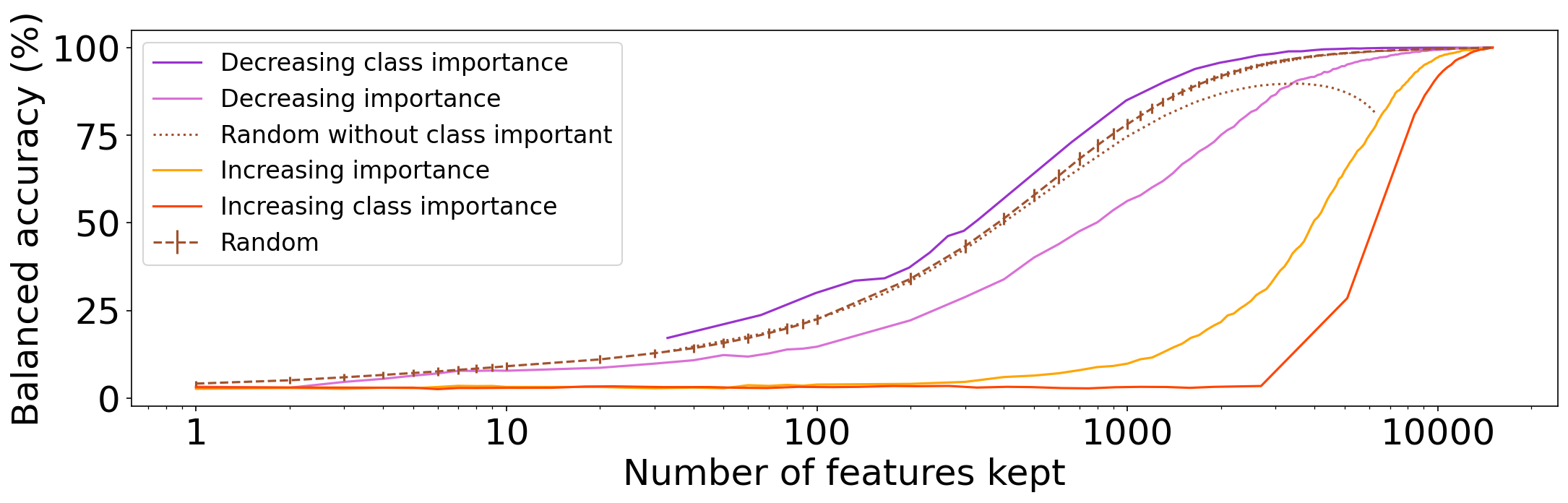}}
\label{fig:simu1}
\end{subfigure}

 \begin{subfigure}[b]{\linewidth}
 \caption{SIMU2}
\centerline{\includegraphics[width=\linewidth]{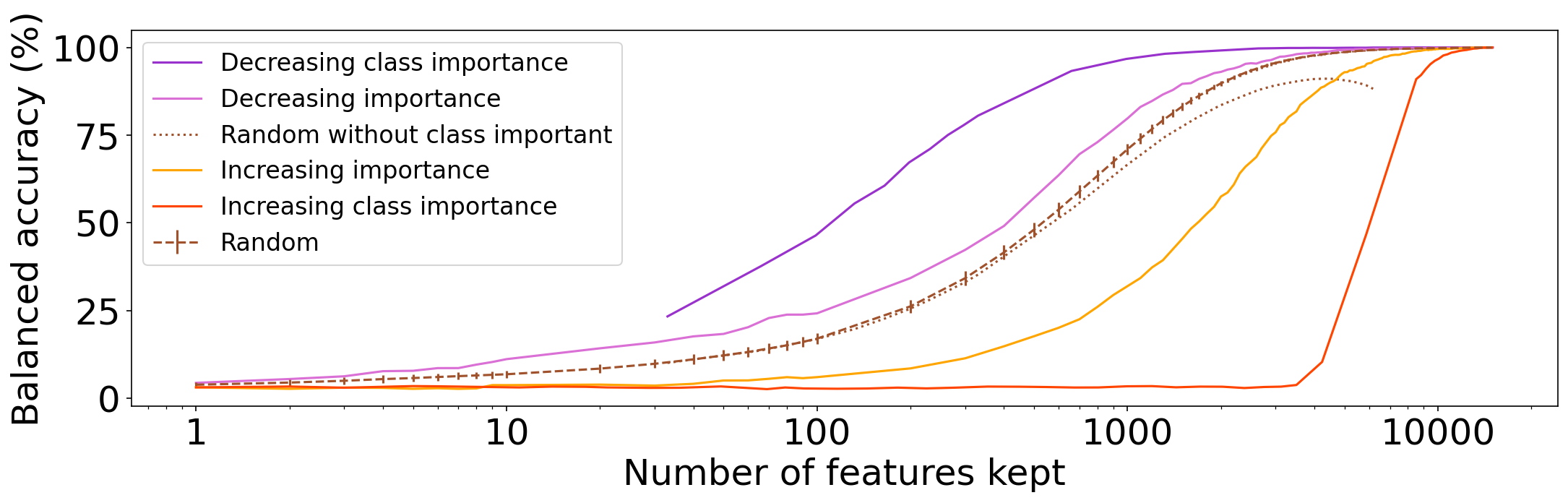}}
\label{fig:simu2}
\end{subfigure}
\caption{Explainability metrics on real (Pan-Can TCGA) and simulated (SIMU1 and SIMU2) data, obtained after learning with logistic regression.}
\label{fig:curves}
\end{figure}

We conduct another experiment to interpret the global meaning of these scores. The balanced accuracy scores obtained from iteratively adding the most important features in the \LR either computed on each class independently (dark purple) or on the whole dataset (light purple) are plot in Fig.~\ref{fig:TCGA}. In the first case, the first point is obtained by keeping the most important gene on average for each of the 33 classes. The curve increases rapidly and contains about the same information than PGI: $\sim 50-100$ genes are enough to perform correctly. The difference between light and dark purple curves highlights that computing the IG score averaged on all sampled (light purple curve) as done classically provide less informative genes that classifying the important genes by classes (dark purple curve).

For comparison, we also computed the balanced accuracy obtained while keeping random features (brown, error bars are standard deviations computed over 100 trials) or random features without the ones identified as important for the classes (brown dots).
The global interpretation of these two curves is that any set of 800 random features, even within the less important ones, is sufficient to get a good performance. The dataset contains highly redundant information, which is not surprising knowing the diversity of cancer cells and the numerous and complex metabolic pathways involving many bio-actors.
These results raise several questions on the interpretation of the explanations: the features proposed as relevant thanks to IG, are not important individually. To go further, we propose to simulate a biologically plausible (yet far more simple) dataset with known ground truth explanations. 

\begin{table}
\centering
\caption{Explainability metrics obtained on the test sets of several datasets, using different supervised learning methods \LR: logistic regression. \MLP: multilayer perceptron. D: diffusion on correlation graph. Metrics are: \PGI: prediction gap on important features. \PGU: prediction gap on unimportant features. \FA: feature agreement.}
\begin{subtable}[h]{\linewidth}
\centering
\caption{Pan-Can TCGA}
\label{tab:TCGA}
\begin{tabular}{ |c|c|c|c|c| } 
 \hline
 Network & LR & MLP & D + LR & D + MLP \\ 
 \hline
 Balanced accuracy ($\uparrow$) & 93.2\% & 94.7\% & 92.5\% & 94.3\%\\
 PGI ($\uparrow$) & 0.9570 & 0.9567 & 0.9750 & 0.9652\\
 PGU ($\downarrow$) & 0.0035 & 0.0197 & 0.0053 & 0.0133\\
 \hline
\end{tabular}
\end{subtable}

\vspace{0.25cm}

\begin{subtable}[h]{\linewidth}
\centering
\caption{Simulations}
\label{tab:Simu}
\begin{tabular}{ |c|c|c|c|c| } 
 \hline
 Dataset & \multicolumn{2}{c|}{SIMU1} & \multicolumn{2}{c|}{SIMU2} \\
 \hline 
 Network & LR & MLP & LR & MLP\\ 
 \hline
 Accuracy ($\uparrow$) & 99.5\%  & 99.5\% & 99.9\%  & 100\%\\
 PGI ($\uparrow$)  & 0.9905  & 0.9714 & 0.9881  & 0.9842\\
 PGU ($\downarrow$) & 0.0007  & 0.0036 & 0.0007  & 0.0039 \\
 FA ($\uparrow$)  & 0.72  & 0.76 & 0.43  & 0.45 \\
 D + FA ($\uparrow$) & 1  & 1 & 0.96 &  1\\
 \hline
\end{tabular}
\end{subtable}

\end{table}

\section{The importance of a mesoscopic scale highlighted on a simulated dataset}
\label{sec:model}

\subsection{A generative model of transcriptomic data,}
For transcriptomic data, various simulation models already exist~\cite{chalise2016intersim, kelly2020graphsim}. Here, we propose to use a generative probabilistic model called \LDA~\cite{blei2003latent} where gene expression is controlled by the activation of metabolic pathways. It is well known for producing documents with a fixed number of words associated with various subjects, yet has already been used in genomics~\cite{yalamanchili2017latent}.
%In RNA-seq, for each sample, the gene expression data is obtained by cutting the DNA in small pieces called reads, duplicating them, sequencing them and finally associating them with a portion of the genome. 
With \LDA, we can generate individual samples (documents) with a fixed number of sequencing reads (words) associated with diverse metabolic pathways (subjects). The expression of the genes in the same pathway appears to be highly correlated in the simulated data.

Formally, the dataset contains expression levels for $G$ genes, themselves grouped in $P$ sets modelling metabolic pathways. An individual sample is described by a set of couples $(\text{gene}, \text{value})$, where the value is a relative number of drawn reads associated with the gene. 
The model requires priors $\boldsymbol{\eta}_p$ on the relative proportion of genes expressed in each pathway $p$ and priors $\boldsymbol{\alpha}_c$ defining the relative proportion of pathways expressed in a sample belonging to the class $c$.

The relative proportion of reads appearing in a pathway is drawn once and for all as $\boldsymbol{\beta}_p \sim \text{Dirichlet}(\boldsymbol{\eta}_p)$. To generate a single sample $s$, belonging to class $c$, two steps are followed:
\begin{enumerate}
    \item Draw the proportion of pathways: $\boldsymbol{\theta}_s~\sim~\text{Dirichlet}(\boldsymbol{\alpha}_c)$,
    \item Drawing of $N$ reads. For each read $i$, 
    \begin{enumerate}
        \item a pathway is assigned $p_{i} \sim \text{Multinomial}(\boldsymbol{\theta}_s)$,
        \item a read is observed $g_i \sim \text{Multinomial}(\boldsymbol{\beta}_{p})$.
    \end{enumerate}
\end{enumerate} 

To make our simulated task more interpretable, the classes are designed to influence a different set of pathways and the pathways to only influence a sparse set of genes.

\subsection{Experimental setting: simulation and learning}
Two simulated datasets, noted as SIMU1 and SIMU2, are generated from the model. They contain 9900 examples with 15000 genes generated from 33 classes. A class is defined by 37 pathways which are over-expressed. By default, all the other pathways have an equal probability to be activated. In SIMU1, the classes have non-overlapping over-expressed pathways.  In SIMU2, pathways can overlap, which more closely reflects the complexity of real signalling pathways. More details on the data generation process are in Appendix~\ref{app:1}. 

The same setting as in Section~\ref{subsec:exp_TCGA} is used to evaluate the quality of the feature-based explanations. As the ground truth is known, \FA can be computed. Additionally, under the intuition that correlated features should have the same importance, we compute a diffused \FA (D + FA) from the IG $\phi_{i}$ diffused by D as defined in Appendix~\ref{app:2}.

\subsection{Interpretation}
We discuss  here in details the results of \LR but similar conclusion can be drawn for all models.
Fig.~\ref{fig:curves} shows the flexibility of the model across two simulations (and can be compared with real \TCGA data).

In Tables~\ref{tab:Simu} the 142 (resp. 178) most important genes should be removed to significantly hurt the prediction for SIMU1 (0.9905 \PGI) (resp. SIMU2 (0.9881 \PGI)). It is consistent with the previous finding as the data is less redundant than in \TCGA by design. \FA shows that among the top-370 genes of a single sample in SIMU1 (number of over-expressed gene per class), 266 genes belong to the over-expressed pathways. In SIMU2, only 43\% of the most important genes belongs to the over-expressed pathways. The metric D + FA obtains better results. This is not surprising given the design of the model where genes inside the same pathway are strongly correlated, and this calls for the seeking of mesoscopic explanations instead of individual feature attributions. Although we did not observe any significant improvement when diffusing the inputs of the model directly (D+LR and D+MLP), it seems promising, from the data perspective, to take into account the correlation graph as an element for the processing of such data, and the diffused feature attribution for explainability.

\section{Conclusion}
\label{sec:ccl}

In this work, we proposed good practices for evaluating biomarkers derived from explainability methods on transcriptomic data. We first evaluated the complexity of a real dataset, \TCGA, by characterising how the accuracy of a network evolves as an increasing proportion of the genes, sorted in different manners, are set to zero. We evaluated two simple metrics: the PGU that allowed to estimate that a specific set of 50 genes are sufficient to correctly classify each sample and the PGI that showed that removing this set was not degrading the prediction. We additionally showed that random subsets of 800 genes were good enough to correctly classify the classes, even when removing the 800 most important genes from the possible choices. These results underline the spread of the information and the ambiguity in defining well behaved explanatory features in gene expression data.
Then, we proposed a simulation tool, based on LDA, with granularity fine enough so that it allowed us to analyse the pertinence of the genes selected by IG.
Interestingly, diffusing the IG score on the correlation matrix led to very strong performance increase in term of explainability ($\sim95- 100 $\% of correctly selected genes).
This interesting direction will be investigated in a future work.

\bibliographystyle{IEEEtran}
\bibliography{references}

\appendix

\subsection{Details on Learning methods and training}
\label{app:2}
Two architectures are used. A \LR with a dropout layer randomly setting to 0 some elements of the input with probability 0.2. A \MLP with 1 hidden layer containing 20 features followed by batch normalization.

In the style of the simple graph convolutional operator~\cite{wu2019simplifying}, we also consider two variations in which the input features are diffused on a graph before applying the \LR (D + \LR) and the \MLP (D + MLP). Intuitively, the value of a feature is smoothed with respect to the values of its neighbours. Here, the graph is a thresholded Pearson correlation graph computed on the training examples. Only positive correlations are kept. Given the graph adjacency matrix $\mat{A}$ and a data sample $\vect{x}$, the diffused features are 
\begin{equation}
    \vect{x}^\text{diffused} = \hat{\mat{D}}^{-\frac{1}{2}}\hat{\mat{A}}\hat{\mat{D}}^{-\frac{1}{2}}\vect{x}\,,
\end{equation}
where $\hat{\mat{A}} = \mat{A} + \mat{I}$ and $\hat{\mat{D}}$ is its diagonal degree matrix. For training, all datasets are split into a training set (60\%) and a test set (40\%).

\subsection{Details on the generation of synthetic datasets}
\label{app:1}
Several priors are used to generate the simulated datasets. Given a pathway $p$, $\boldsymbol{\alpha}_c[p] = 8$ if it is over-expressed in class $c$ and $\boldsymbol{\alpha}_c[p] = 2$ otherwise. Given a gene $g$, $\boldsymbol{\eta}_p[g] = 5$ if gene $g$ is expressed in pathway $p$ and $\boldsymbol{\eta}_p[g] = 0$ otherwise. In SIMU1 and SIMU2, 37 pathways are over-expressed per class, that is 2.5\% of the genes. In SIMU1, there are 1500 pathways in total. 
In SIMU2, a random subset of genes is assigned to each pathway. In practice, a Bernoulli variable of parameter $\frac{1}{1500}$ is drawn for each gene (10 genes per pathway in average). If the result is 1, the gene is assigned to the pathway. To avoid  having a too small number of genes, the process is run again if the number of assigned genes is less than 3. To avoid a large number of genes associated with no pathway (null expression), 3000 pathways are generated. 

The proportion of reads resulting from a pathway $p$ is $\boldsymbol{\beta}_p \sim \text{Dirichlet}(\boldsymbol{\eta}_p)$. To study the ability to retrieve the perturbed pathways, we make sure that if a gene $g$ is associated with a perturbed pathway $p$, we have $\boldsymbol{\beta}_p[g] \geq 0.01$.

\subsection{Details on the explainability procedure}
\label{app:3}
The attribution scores are computed only on the correctly classified samples of the test set. As the scale of the most important attributions varies significantly w.r.t.  samples, we scale the norm of the importance scores of each sample to~1 for their comparison. 

\end{document}